\documentclass[a4paper,12pt]{article}
\usepackage{amsmath,amsfonts,amssymb,cite}

\begin{document}

\begin{center}
{\LARGE\bf Low-energy holographic models for QCD}
\end{center}

\begin{center}
{\large S. S. Afonin}
\end{center}

\begin{center}
{\it V. A. Fock Department of Theoretical Physics, Saint-Petersburg
State University, 1 ul. Ulyanovskaya, 198504, Russia
}
\end{center}

\begin{abstract}
We consider the bottom-up holographic models for QCD which contain
the ultraviolet (UV) cutoff. Such models are supposed to describe
exclusively the low-energy sector of QCD. The introduction of UV
cutoff in the soft wall model is shown to result in a model with
qualitatively different predictions. The ensuing model seems to be
able to incorporate the constituent quark mass. It is also
demonstrated that in order to reproduce the results of the usual
soft wall model for the vector and higher spin mesons in the
presence of the UV cutoff one can consider the flat bulk space with
a modified dilaton background.
\end{abstract}

\section{Introduction}

The fundamental theory of strong interactions --- Quantum
Chromodynamics (QCD) --- is known to be highly difficult for the
analytical analysis at low energies because of the strong coupling
regime. Meanwhile this regime triggers the most interesting
phenomena in QCD and therefore the low-energy domain is of extreme
theoretical interest. For this reason, it has become a common
practice to replace the low-energy QCD by some effective
description. Examples of such descriptions include the Sigma-model,
Nambu--Jona-Lasinio model, chiral quark model and others. They
provide a simple framework for the analysis of some aspects of
non-perturbative QCD and lead to relations which are qualitatively
or even semiquantitatively right. In spite of much efforts, no any
rigorous relation of those models to QCD has been established.
Nevertheless the simplicity of theoretical setup makes the effective
models very interesting and useful approach to the phenomenology of
strong interactions.

Recently a qualitatively new approach emerged on the market that
intends to replace QCD at low and intermediate energies and perhaps
to do more than the traditional effective models. This approach was
inspired by the ideas of gauge/gravity correspondence from the
string theory~\cite{mald,witten} which suggest that the 4D strongly
coupled gauge theories may have a dual semiclassical description in
the 5D anti-de Sitter (AdS) space. There is no recipe how to
construct the holographic duals for the confining theories like QCD.
One can try however to guess such a dual model for a limited class
of physical problems. This ambitious idea received a concrete
realization in the form of so-called bottom-up holographic
models proposed some time ago~\cite{son1,pom}. They have been
extensively applied to various problems more or less successfully.
On the other hand, the bottom-up approach faced a certain criticism,
one of worrying point is that the ensuing models are usually matched
to QCD in the UV regime where QCD represents a weakly coupled theory
due to its asymptotic freedom. At the same time, one expects that
the corresponding dual theory (if it exists) should be then in the
strongly coupled regime, hence, the applicability of semiclassical
treatment for the latter becomes questionable. It is therefore
interesting to construct bottom-up models which are free of this
conceptual drawback. A straightforward idea consists in imposing the
UV cutoff that removes the high-energy region from a tentative
holographic model. In the present Letter, we undertake an
exploratory consideration of two bottom-up models with the UV
cutoff\footnote{A similar issue was considered numerically in Ref.~\cite{evans}.}.

The paper is organized as follows. In Section~2, we introduce the UV
cutoff to a simple soft wall model~\cite{son2} and discuss the
resulting model which turns out to be quite different. The main
results of the soft wall model may be however reproduced if one
accepts the flat bulk space after imposing the UV cutoff. This is
shown in Section~3. Our conclusions are summarized in Section~4.

\section{Soft wall model with the UV cutoff}

A straightforward possibility in constructing the holographic models
for the IR domain of QCD seems to impose the UV cutoff in the
existing bottom-up holographic models. As an educative exercise let
us analyze how the Soft Wall (SW) model introduced in
Ref.~\cite{son2} is modified after imposing the UV boundary
$z_{\text{UV}}$ that corresponds to the inverse scale of the onset
of the strongly coupled regime, about 1~GeV$^{-1}$. The simplest
version of the SW model is defined by the action~\cite{son2}
\begin{equation}
\label{1}
S=-\frac{1}{4g_5^2}\int d^4\!x\, dz \sqrt{g}\,
e^{-\Lambda^2z^2}F_{MN}F^{MN},
\end{equation}
where $F_{MN}=\partial_MV_N-\partial_NV_M$, $M,N=0,1,2,3,4$, and the
metric of the AdS$_5$ space having radius $R$ is given by
($\mu=0,1,2,3$)
\begin{equation}
\label{2} ds^2=\frac{R^2}{z^2}\left(dx_{\mu}^2-dz^2\right); \quad
0\leq z<\infty.
\end{equation}
The dilaton background $e^{-\Lambda^2z^2}$ introduces the scale
$\Lambda$ that determines the slope of the linear mass spectrum of
normalizable modes, $m_n^2=4\Lambda^2(n+1)$, $n=0,1,2,\dots$.

Without loss of generality we identify the UV cutoff $z_{\text{UV}}$
with the radius of the AdS$_5$ space, $z_{\text{UV}}=R$. It means
that in performing the integral over $z$,
\begin{equation}
\label{3}
\int_0^{\infty}dz=\int_0^Rdz+\int_R^{\infty}dz,
\end{equation}
we do not consider the region $0\leq z<R$ since we do not expect the
validity of the semiclassical approximation in that region. In the
axial gauge, $V_z(x,z)=0$, the equation of motion for the 4D Fourier
transform $V_{\mu}(q,z)$ of the transverse components,
$\partial_{\mu}V^{\mu}(x,z)=0$, takes the form
\begin{equation}
\label{4}
-\partial_z\left(\frac{e^{-\Lambda^2z^2}}{z}\partial_z
V_{\mu}(q,z)\right)=q^2\frac{e^{-\Lambda^2z^2}}{z}V_{\mu}(q,z).
\end{equation}
Letting $V_{\mu}(q,z)=v(q,z)V_{\mu}^0(q)$, we require that
$v(q,R)=1$; the source for the 4D vector current in the momentum
space is then given by $V_{\mu}^0(q)$. The corresponding solution to
the Eq.~\eqref{4} bounded as $z\rightarrow\infty$ is
\begin{equation}
\label{5}
v(q,z)=\frac{U(-q^2/4\Lambda^2,0,\Lambda^2z^2)}{U(-q^2/4\Lambda^2,0,\Lambda^2R^2)},
\end{equation}
where $U$ is the Tricomi confluent hypergeometric function.

Evaluating the action~\eqref{1} on the solution leaves the boundary
term
\begin{equation}
\label{6}
S=-\frac{R}{2g_5^2}\int
d^4\!x\left(\frac{e^{-\Lambda^2z^2}}{z}V_{\mu}\partial_z
V^{\mu}\right)_{z=R}.
\end{equation}
According to the conjecture of the AdS/CFT
correspondence~\cite{witten}, the vector two-point correlation
function,
\begin{equation}
\label{7} \int d^4\!x\, e^{iqx}\left\langle
J_{\mu}(x)J_{\nu}(0)\right\rangle=(q_{\mu}q_{\nu}-q^2\eta_{\mu\nu})\Pi_V(Q^2);\quad
Q^2=-q^2,
\end{equation}
is given by the second derivative of the term~\eqref{6} with respect
to the source $V_{\mu}^0$,
\begin{equation}
\label{8} \Pi_V(Q^2)=-\frac{R
e^{-\Lambda^2z^2}}{g_5^2Q^2}\left.\frac{v\partial_z
v}{z}\right|_{z=R}.
\end{equation}
Using the normalization $v(q,R)=1$ and the property $\partial_x
U(a,0,x)=-a U(1+a,1,x)$, we obtain finally
\begin{equation}
\label{9}
\Pi_V(Q^2)=\frac{Re^{-\Lambda^2R^2}}{2g_5^2}\frac{U(1+Q^2/4\Lambda^2,1,\Lambda^2R^2)}{U(Q^2/4\Lambda^2,0,\Lambda^2R^2)}.
\end{equation}
The expression~\eqref{9} has poles $q_n^2=4\Lambda^2f_n$,
$n=0,1,2,\dots$, where $f_n\rightarrow n+1$ as $\Lambda
R\rightarrow0$. At $\Lambda R>0$, $f_n$ tends to the equidistant
behavior, $f_{n+1}-f_n\rightarrow1$, at large $n$. For instance, the
choice $\Lambda R=1$ leads to
$f_n\approx1.57,2.84,4.05,5.22,6.37,\dots$. In contrast to the SW
model, the residues are vanishing as $n\rightarrow\infty$. In the
limit $\Lambda R\rightarrow0$, the residues of the expression
$\frac{U(1+Q^2/4\Lambda^2,1,\Lambda^2R^2)}{U(Q^2/4\Lambda^2,0,\Lambda^2R^2)}$
tend to $4\Lambda^2$. Thus, taking into account the general factor in the
relation~\eqref{9}, we arrive at the same residues as in the vector
correlator of the SW model,
\begin{equation}
\label{9b}
\Pi_V^{(\text{SW})}(Q^2)=\frac{R}{2g_5^2}\left[\sum_{n=0}^{\infty}\frac{4\Lambda^2}{Q^2+4\Lambda^2(n+1)}+\gamma-
\sum_{k=1}^{\infty}\frac{1}{k}-\log(z^2\Lambda^2)\right]_{z\rightarrow0}
\end{equation}
The expression~\eqref{9b} contains two infinite terms since the
limit $\Lambda R=0$ is implied from the very beginning. They are
subtracted in the final answer while in our case we did not make any
subtractions.

The introduction of the UV cutoff may solve the problem with a
natural description of the Chiral Symmetry Breaking (CSB) within the
SW model. We remind the reader the essence of the problem. The
simplest way for incorporating the CSB consists in introducing an
action quadratic in a scalar field $X$~\cite{son1,pom},
\begin{equation}
\label{10}
S_{\text{CSB}}=\int d^4\!x\, dz \sqrt{g}\,
e^{-\Lambda^2z^2}\left(|\partial_M X|^2+\frac{3}{R^2}|X|^2\right),
\end{equation}
where the background field $\frac{2}{z}X(z)$ corresponds to the
quark bilinear operator $\bar{q}q$ and this correspondence dictates
the mass term from the relation~\eqref{25}. In the full model, the
usual derivative in~\eqref{10} should be replaced by the covariant
one to have a coupling of $X(z)$ with the vector fields; this is not
relevant for the present discussion.

According to the AdS/CFT based prescriptions, the bulk scalar field
responsible for the CSB should have the following UV
asymptotics~\cite{kleb}
\begin{equation}
\label{13}
X(z)_{z\rightarrow0}\sim Mz+\Sigma z^3,
\end{equation}
the coefficient $M$ represents then the quark mass and $\Sigma$ is
the chiral condensate. However, the equation of motion for $X$
following from the action~\eqref{10} has only one solution bounded
as $z\rightarrow\infty$, $X(z)=zU(\frac12,0,\Lambda^2z^2)$. Its
expansion around $z=0$ reads
\begin{equation}
\label{14}
X(z)_{z\rightarrow0}\sim 2\Lambda
z+\left(1+\gamma-\log{4}+\log(\Lambda^2z^2)\right)\Lambda^3z^3,
\end{equation}
where $\gamma\approx0.577$ is the Euler constant. Thus, $\Sigma$ is
proportional to $M$. Since this is not what one expects in QCD, the
given description of the CSB was rejected in Ref.~\cite{son2}.
However, if we assume that the prescription~\eqref{13} holds
approximately also for a non-zero $z=R$ and take into account that
the model now is defined in the IR domain only, $z\geq R$, the
behavior above is exactly what one expects in the effective
description of the low-energy QCD --- the light quarks acquire a
constituent (called also dynamical) mass, $M\approx320$~MeV, that is
proportional to $\Sigma$. For example, within the
Nambu--Jona-Lasinio model for the low-energy QCD~\cite{klev}, the
relation is $M=-2GN_f\Sigma+M_0$, where $G$ represents the
four-fermion coupling and $M_0$ is the current quark mass.

We may attempt to exploit the relation~\eqref{14} for a rough
estimate of the quantity $\Lambda R$. Multiplying~\eqref{14} by a
constant $C$ and comparing with~\eqref{13} one obtains
$C=\frac{M}{2\Lambda}$ and
\begin{equation}
\label{15}
\Sigma\simeq\frac12\left(1+\gamma-\log{4}+\log(\Lambda^2R^2)\right)M\Lambda^2.
\end{equation}
Taking $\Lambda=550$~MeV from the approximate fits for the slope
$4\Lambda^2$ in the vector mass spectrum and
$\Sigma=(-235\,\text{MeV})^3$, we have the estimate $\Lambda
R\approx0.8$, i.e. $R\approx\frac{1}{0.7\,\text{GeV}}$. This means
that the model is defined roughly below the mass of the
$\rho$-meson.

It is interesting to note a similarity between the
expression~\eqref{15} and the relation for the chiral condensate
provided by the Nambu--Jona-Lasinio model regularized by the 4D
momentum cutoff $\Lambda_{\text{cut}}$~\cite{klev},
$\Sigma\sim-M\Lambda_{\text{cut}}^2+\mathcal{O}(M^3)$.

\section{Flattened soft wall model}

In building the holographic models that describe exclusively the
low-energy sector of QCD we do not need to impose the AdS$_5$ metric
in the UV limit since the conformal symmetry is believed to be
strongly broken\footnote{There are, however, some suggestions in the
literature (see, e.g.,~\cite{brod}) that the conformal symmetry is
restored at very low energies.}. Phenomenologically the most
satisfactory bottom-up model seems to be the SW one. An interesting
question arises: Which holographic models with the UV cutoff
reproduce the basic results of the SW model, say the form of the
vector two-point correlator? We are going to argue that such a model
can be constructed in the flat 5D space.

\subsection{The model}

The metric~\eqref{2} can be cast into the form
\begin{equation}
\label{16}
ds^2=e^{-ky/R}dx_{\mu}^2-dy^2;\quad k=2,
\end{equation}
where $y=R\log\frac{z}{R}$ and $-\infty<y<\infty$. The fifth
coordinate $y$ has then the physical meaning of the logarithm of
energy scale. The UV bounded interval $R\leq z<\infty$ translates
into $0\leq y<\infty$. Let us consider the case of the flat space,
$k=0$ in~\eqref{16}, and write an action for the free vector field
with some unknown dilaton background $f(y)$,
\begin{equation}
\label{17}
S=-\frac{1}{4}\int d^4\!x\, dy\, e^{-f(y)}F_{MN}F^{MN}.
\end{equation}
Proceeding as in Section~2, we arrive at the equation of motion for
the field $v(q,y)$,
\begin{equation}
\label{18} -\partial_y\left(e^{-f}\partial_yv\right)=q^2e^{-f}v.
\end{equation}
The substitution
\begin{equation}
\label{19}
\varphi=e^{f/2}\psi,
\end{equation}
transforms the Eq.~\eqref{18} into a Schr\"{o}dinger like equation
\begin{equation}
\label{20}
-\psi''+\left(\frac{(f')^2}{4}-\frac{f''}{2}\right)\psi=q^2\psi.
\end{equation}
Following the Ref.~\cite{son2} we try the ansatz ($R$ below is
simply a constant of dimension $[\text{mass}^{-1}]$)
\begin{equation}
\label{21} f=Ay^2+B\log\frac{y}{R}
\end{equation}
and consider the exactly solvable case $A=\Lambda^2$, $B=1$. The
"potential" of Eq.~\eqref{20} is then
\begin{equation}
\label{22}
\frac{(f')^2}{4}-\frac{f''}{2}=\Lambda^4y^2+\frac{3}{4y^2}.
\end{equation}
The further results are identical to those of the SW model but with
$z$ replaced by $y$: The normalizable solutions form a discrete set
of modes with the masses $q_n^2=m_n^2$,
\begin{equation}
\label{22b}
m_n^2=4\Lambda^2(n+1);\quad n=0,1,2,\dots,
\end{equation}
and wave functions
\begin{equation}
v(y)=\sqrt{\frac{2n!}{(n+1)!}}\Lambda^2y^2L_n^1(\Lambda^2y^2),
\end{equation}
where $L_n^1$ are associated Laguerre polynomials. The calculation
of the vector two-point correlator yields also the same result,
$\Pi_V(Q^2)=-\frac12\psi(1+\frac{Q^2}{4\Lambda^2})+\text{\it
const}$, where the digamma function $\psi$ has poles located at
$q_n^2=m_n^2$ given by~\eqref{22b}.

Thus we see that without the CSB the SW model and the flattened SW
model with appropriately chosen background lead to identical
predictions for the vector mesons. The appearance of factor $y^{-1}$
in the dilaton background may be interpreted as an "encoded" rest of
conformal symmetry ( = AdS metric). The UV limit, $y\rightarrow0$,
of the flattened SW model, however, does not correspond to the UV
limit in QCD and can be taken about 1~GeV in order to describe the
strongly coupled regime only. The description of the CSB requires an
analogue of the prescription~\eqref{13} for the flat space. This
problem will be considered somewhere.

\subsection{Higher spin fields}

In what follows we will try to include the Higher Spin Fields (HSF)
into the Flattened Soft Wall (FSW) model. The free massless HSF are
described by symmetric double traceless tensors $\Phi_{M_1\dots
M_J}$~\cite{frons}. The corresponding action is invariant under the
gauge transformations $\delta\Phi_{M_1\dots
M_J}=\nabla_{(M_1}\xi_{M_2\dots M_J)}$, where $\nabla$ is covariant
derivative with respect to the general coordinate transformations
and the gauge parameter $\xi$ represents a traceless symmetric
tensor. We first remind the reader the result of incorporation of
HSF into the usual SW model~\cite{son2}. The quadratic part of the
action reads
\begin{equation}
\label{23} S^{(J)}=\frac{1}{2}\int d^4\!x\, dz \sqrt{g}\,
e^{-\Lambda^2z^2}\left(\nabla_N\Phi_{M_1\dots
M_J}\nabla^N\Phi^{M_1\dots M_J}+\dots\right),
\end{equation}
where further terms are omitted. As was argued in Ref.~\cite{katz},
in the axial gauge, $\Phi_{z\dots}=0$, the action for a rescaled
field $\Phi=\left(\frac{z}{R}\right)^{2(1-J)}\tilde{\Phi}$ contains
only the first kinetic term written in~\eqref{23}. The resulting
equation of motion for $\tilde{\Phi}(x)$ in the SW model results in
the mass spectrum~\cite{son2}
\begin{equation}
\label{24}
m_{n,J}^2=4\Lambda^2(n+J),
\end{equation}
which generalizes the spectrum for the $J=1$ case. The
spectrum~\eqref{24} corresponds to the poles of Veneziano amplitude
and is expected (up to some additional intercept) in the effective
string description of QCD. It is interesting to note that such a
spectrum seems to hold approximately in the phenomenology of light
mesons~\cite{a1} (again up to a general shift). For the universality
of intercept, however, one must replace $J$ by the relative angular
momentum $L$ of pions produced via the strong decay of resonance
under consideration~\cite{a2}. In the framework of the
non-relativistic quark model, the account for the quark spin leads
to the relation $J=L, L\pm1$.

In order to include the HSF into the FSW model we first reinterpret
the corresponding result of the Ref.~\cite{son2}. The rescaled field
$\tilde{\Phi}_{M_1\dots M_J}$ corresponds to a twist-two operator
with the canonical dimension $\Delta=J+2$~\cite{son2}. According to
the AdS/CFT prescriptions~\cite{witten}, the masses of 5D fields
propagating in the AdS$_5$ space which correspond to the $p$-form
operators of dimension $\Delta$ in the equivalent 4D theory are
given by
\begin{equation}
\label{25}
R^2m_5^2(p)=(\Delta-p)(\Delta+p-4),
\end{equation}
The question arises how this relation could show up in the
holographic models describing the $J>1$ mesons? It is easy to
observe that the action~\eqref{23} in terms of the rescaled field
$\tilde{\Phi}$ can be written as
\begin{equation}
\label{26}
S^{(J)}=\frac{1}{2}\int d^4\!x\, dz
\sqrt{g}\left(\frac{R}{z}\right)^{R^2m_5^2(J)}\!
e^{-\Lambda^2z^2}\nabla_N\tilde{\Phi}_{M_1\dots
M_J}\nabla^N\tilde{\Phi}^{M_1\dots M_J}.
\end{equation}
This implies that the description of the HSF proposed
in~\cite{katz,son2} is equivalent to the assumption that such fields
couple to different 5D backgrounds. The form of these backgrounds is
dictated by the relation~\eqref{25} in which $p$ is replaced by $J$.
Contracting the Lorentz indices in~\eqref{26} for the case of pure
AdS$_5$ space~\eqref{2} one arrives finally at the action
\begin{equation}
\label{27}
S^{(J)}=\frac{1}{2}\int d^4\!x\, dz
\left(\frac{R}{z}\right)^{2J-1}\!
e^{-\Lambda^2z^2}(\nabla_N\tilde{\Phi})^2.
\end{equation}
In the flat space we expect that the incorporation of the HSF should
follow the same principle. It is easy to verify that postulating in
the axial gauge the action~\eqref{27} for the flat space (with $z$
replaced by $y$) leads to the spectrum~\eqref{24} which represents
the eigenvalues of the equation of motion,
\begin{equation}
\partial_y\left(\frac{\partial_y\tilde{\Phi}_n}{y^{2J-1}}\right)+\frac{m^2_n}{y^{2J-1}}\tilde{\Phi}_n=0,
\end{equation}
for the 4D Fourier transform $\tilde{\Phi}(q^2,y)$, $q^2_n=m^2_n$.
The 5D background in the action~\eqref{27} generalizes that of the
previous Section to the $J>1$ states. The parameter $R$
in~\eqref{27} can be regarded just as a dimensional parameter, say
$R=1/\Lambda$.



\section{Conclusions}

We have considered a couple of bottom-up holographic models for
QCD with the UV cutoff imposed. They are closer in spirit to the
traditional effective models for the low-energy QCD which possess
in a direct or indirect way the UV cutoff showing the
applicability of a model. Like in the effective field theories,
the present approach neglects the running of some physical
quantities with the energy scale.

In the first part, the UV cutoff was introduced to the
conventional soft wall model and it was demonstrated that the
ensuing model differs substantially from the original one. We also
argued that the disadvantage of the original model in describing
the chiral symmetry breaking --- the resulting proportionality of
the quark mass to the chiral condensate --- may be converted into
an advantage after imposing the UV cutoff as exactly this pattern
holds in the low-energy effective approaches.

In the second part, we put forward a kind of soft wall model in the
flat 5D space motivating this choice by the absence of conformal
invariance at low energies where the model is applicable. We
proposed an ansatz for modification of dilaton background that
reproduces the predictions of the soft wall model both for the
vector mesons (neglecting the chiral symmetry breaking) and for the
higher spin mesons. The description of the chiral symmetry breaking
within such a setup is left for the future.

\section*{Acknowledgments}

The work is supported by RFBR, grant 09-02-00073-a, by the Dynasty Foundation and by the
Alexander von Humboldt Foundation.

\end{document}